\def\e{{\rm e}}
\def\del{\partial}

\def\vev#1{\langle #1 \rangle}

\def\del{\partial}
\def\dslash{\del\kern-0.55em\raise 0.14ex\hbox{/}}

\def\rough#1{\raise.3ex\hbox{$#1$\kern-.75em\lower1ex\hbox{$\sim$}}}

\newcommand{\PRD}[3]{{\it Phys. Rev.} {\bf D{#1}} (19{#3}) {#2}}
\newcommand{\PRDM}[3]{{\it Phys. Rev.} {\bf D{#1}} (20{#3}) {#2}}

\newcommand{\NPB}[3]{{\it Nucl. Phys.} {\bf B{#1}} {#2} (19{#3})}
\newcommand{\NPBM}[3]{{\it Nucl. Phys.} {\bf B{#1}} (20{#2}) {#3}}
\newcommand{\PLB}[3]{{\it Phys. Lett.} {\bf {#1}B} (19{#3}) {#2}}
\newcommand{\PLBM}[3]{{\it Phys. Lett.} {\bf {#1}B} (20{#3}) {#2}}

\newcommand{\PTP}[3]{{\it Prog. Theor. Phys.} {\bf {#1}} (19{#3}) {#2}}

\newcommand{\ANN}[3]{{\it Ann. Phys. (N.Y.)} {\bf {#1}}, {#2} (19{#3})}

\newcommand{\MPL}[3]{{\it Mod. Phys. Lett.} {\bf A{#1}} (19{#3}) {#2}}
\newcommand{\MPLM}[3]{{\it Mod. Phys. Lett.} {\bf A{#1}} (20{#3}) {#2}}

\newcommand{\jhep}[3]{{\it J. High Energy Phys.}{\bf {#1}}, {#2} (20{#3})}

\newcommand{\hepph}[1]{{\tt hep-ph/#1}}

\newcommand{\hmu}{\hat\mu}

\documentclass[12pt]{article}
\usepackage{graphicx}
\begin{document}
\baselineskip=18pt
\begin{titlepage}
\begin{flushright}
KUNS-1940\\
OU-HET-495/2004
\end{flushright}
\begin{center}{\Large\bf Partial Gauge Symmetry Breaking via Bare Mass}
\end{center}
\vspace{1cm}
\begin{center}
Naoyuki {Haba}$^{(a),}$
\footnote{E-mail: haba@ias.tokushima-u.ac.jp}
Kazunori {Takenaga}$^{(b),}$
\footnote{E-mail: takenaga@het.phys.sci.osaka-u.ac.jp}
Toshifumi {Yamashita}$^{(c),}$
\footnote{E-mail: yamasita@gauge.scphys.kyoto-u.ac.jp},
\end{center}
\vspace{0.2cm}
\begin{center}
${}^{(a)}$ {\it Institute of Theoretical Physics, University of
Tokushima, Tokushima 770-8502, Japan}
\\[0.2cm]
${}^{(b)}$ {\it Department of Physics, Osaka University, 
Toyonaka, Osaka 560-0043, Japan}
\\[0.2cm]
${}^{(c)}${\it Department of Physics, Kyoto University, Kyoto, 
606-8502, Japan}
\end{center}
\vspace{1cm}
\begin{abstract}
We study gauge symmetry breaking patterns in supersymmetric
gauge models defined on $M^4\times S^1$.
Instead of utilizing the Scherk-Schwarz mechanism, supersymmetry is 
broken by bare mass terms for gaugino and squarks.
Though the matter content is the same, depending on 
the magnitude of the bare mass, the gauge symmetry
breaking patterns are different. We present two examples, in one of
which the partial gauge 
symmetry breaking $SU(3)\rightarrow SU(2)\times U(1)$ is realized.
\end{abstract}
\end{titlepage}
\newpage
\section{Introduction}
Gauge symmetry breaking through the Wilson line 
phases (Hosotani mechanism)\cite{hosotania}
is one of the most important dynamical phenomena when one considers
physics with extra
dimensions. Component gauge fields for compactified directions, which
are closely related with the Wilson line phases, become dynamical 
variables and develop vacuum expectation values
to break dynamically gauge symmetry at the quantum level.
\par
It is expected that the mechanism is crucial for the scenario
of the gauge-Higgs 
unification \cite{gaugehiggs1}-\cite{hoso}, in which 
the component 
gauge field for the compactified direction plays 
the role of ``Higgs'' scalars. The Higgs scalar is unified in 
part of the gauge potential by the gauge principle, so that the 
arbitrariness associated with the Higgs sector of the standard
model can be resolved.
\par
The mechanism induces dynamical gauge symmetry breaking, and it
is important to study how the gauge symmetry is broken and/or what
the gauge symmetry breaking patterns depend on. The gauge symmetry 
breaking pattern has been studied extensively in 
many models \cite{models}. It has been reported that 
the gauge symmetry breaking patterns depend on matter content, that is, 
the number of massless particles and their boundary conditions. 
\par
One of the authors (K.T) studied effects of bare mass on the 
Hosotani mechanism on the ground that how global quantities like the
Wilson line phase is affected by massive 
particles \cite{takenaga}. It has been found 
that the existence of the bare mass, depending on its magnitude,
actually changes the vacuum structure of the theory, so that the 
gauge symmetry breaking patterns are modified and are different from that
in the theory with only massless particles.  
\par
In this paper we generalize the previous work \cite{takenaga} to
a higher rank gauge
group $SU(3)$ and study the gauge symmetry breaking patterns when one
considers matter fields with bare mass terms. 
We study how the bare mass affects the gauge symmetry breaking patterns.
Namely, we are interested 
in whether or not the partial gauge symmetry breaking
$SU(3)\rightarrow SU(2)\times U(1)$ can be realized in the model like 
it happened in the model with only massless 
particles \cite{hatanaka}. We will show two 
examples, one of which does not
realizes the partial gauge symmetry breaking, while the other of
which actually does the desirable partial gauge symmetry breaking. 
In the two models the matter content is the same to each 
other, but the relative magnitude among the bare masses is 
chosen to be different. 
\par
We study ${\cal N}=1$ supersymmetric gauge models in five 
dimensions with
one space coordinate being compactified on $S^1$. And we
introduce matter fields belonging to the 
adjoint or fundamental representations 
under the gauge group. The matter fields possess gauge invariant but
non-supersymmetric bare 
mass terms. One usually resorts to the Scherk-Schwarz 
mechanism \cite{ss} 
in order to break supersymmetry, which gives a natural
and simple framework to discuss the gauge symmetry breaking
through the Wilson lines 
in supersymmetric gauge models \cite{takenagab}. In this paper,
we shall consider the non-supersymmetric bare mass term instead of
the Scherk-Schwarz mechanism and see how the term affects on the gauge
symmetry breaking patterns. 
\section{Effective potential of models}
Let us consider an ${\cal N}=1$ supersymmetric gauge theory in
five dimensions. 
The on-shell degrees of freedom of the vectormultiplet are a (Dirac) 
gaugino $(\lambda_D)$ \footnote{$n$ Dirac spinors are
equivalent to $2n$ symplectic (pseudo) Majorana spinors. $\lambda_D$ can
be decomposed into Majorana spinors $\lambda, \lambda'$ in four
dimensions.}, a real scalar $(\Sigma)$ and the gauge field
$(A_{\hmu})$. All the fields belong
to the adjoint representation under the gauge group. We also introduce 
hypermultiplets belonging to the adjoint or fundamental 
representation under the gauge group. The on-shell degrees of freedom
of the hypermultiplet 
consist of a Dirac fermion $(\psi_D)$ and two complex 
scalars $(\phi_{i=1, 2})$.
\par
We study the model on $M^4\times S^1$, where $M^4$ and $S^1$ are
the four-dimensional Minkowski space and a circle, respectively.
As is well known, the zero mode of the component gauge field 
for the $S^1$ direction $A_y$ can develop 
vacuum expectation values. We parameterize $\vev{A_y}$ as 
\begin{equation}
gL\vev{A_y}=\mbox{diag}(\theta_1, \theta_2, \cdots, \theta_N)
\qquad\mbox{with}\qquad \sum_{i=1}^{N}\theta_i=0
\label{background}
\end{equation}
for $SU(N)$ gauge group. Here $g$ is the gauge 
coupling constant in five 
dimensions and $L(=2\pi R)$ stands for the length of the circumference 
of $S^1$. In order to study
the vacuum structure of the model, one usually 
evaluates the effective potential for the phases $\theta_i$'s in one-loop 
approximation by expanding 
the fields around the background (\ref{background}). 
The phase is determined dynamically by minimizing the
effective potential and 
the gauge symmetry can be broken down.
\par
If the theory has exact supersymmetry, the effective potential 
vanishes due
to the cancellation of the quantum correction to the background
between bosons and fermions. One needs to break supersymmetry 
in order
to obtain a nonvanishing effective potential. The Scherk-Schwarz
mechanism of supersymmetry breaking provides us a natural 
and simple
framework to obtain the nonvanishing effective potential. 
Here we introduce bare mass terms in such a way that 
the term breaks supersymmetry explicitly even if we do not
consider the Scherk-Schwarz mechanism.
\par
Following the standard prescription we obtain the effective
potential for the $SU(N)$ gauge group in one-loop 
approximation \cite{pomarol}, up to $\theta_i$-independent terms, 
\begin{eqnarray}
{\bar V}_{eff}
&=& V_{eff}/({{3\over {4\pi^2 L^5}}})\nonumber \\
&=&(-4)\sum_{n=1}^{\infty}\sum_{i, j=1}^{N}{1\over n^5}
\cos[n(\theta_i -\theta_j)]
+4\sum_{n=1}^{\infty}\sum_{i, j=1}^{N}{1\over n^5}D(z_g,n)
\cos[n(\theta_i -\theta_j-\beta)]\nonumber\\
&+&
4N^{adj}\sum_{n=1}^{\infty}\sum_{i, j=1}^{N}{1\over n^5}
\cos[n(\theta_i -\theta_j)]
-4N^{adj}\sum_{n=1}^{\infty}\sum_{i, j=1}^{N}{1\over n^5}
D(z_{adj},n)
\cos[n(\theta_i -\theta_j-\beta)]\nonumber\\
&+&
(2\times 4N^{fd})\sum_{n=1}^{\infty}\sum_{i=1}^{N}{1\over n^5}
\cos[n(\theta_i)]-(2\times 4N^{fd})
\sum_{n=1}^{\infty}\sum_{i=1}^{N}{1\over n^5}
D(z_{fd},n)\cos[n(\theta_i-\beta)],\nonumber\\
\end{eqnarray}
where 
\begin{equation}
D(z_i, n)\equiv \left(1 + z_in + {{(z_in)^2}\over 3}\right)
\e^{-nz_i},\qquad (i=g,~~adj,~~fd).
\label{bessel}
\end{equation}
The bare mass for the gaugino and squark belonging to the fundamental
(adjoint) representation is denoted 
by $m_{g}$ and $m_{fd}(m_{adj})$, respectively, and we have defined the 
dimensionless quantity $z_i\equiv m_iL$. We have introduced the
common bare mass 
for $\phi_1$ and $\phi_2$ for simplicity and have taken the
positive mass terms for them in order to avoid the instability 
of the system. $N_{fd(adj)}$ is the number of 
flavors for the fundamental (adjoint) hypermultiplet.
The phase $\beta$ comes from the Scherk-Schwarz mechanism
of supersymmetry breaking, 
\begin{equation}
\left(
\begin{array}{c}
\lambda\\
\lambda'
\end{array}\right)
(y+L)=\e^{i\beta\sigma_3}
\left(
\begin{array}{c}
\lambda\\
\lambda'
\end{array}\right)
(y),\qquad 
\left(
\begin{array}{c}
\phi_1^{fd(adj)}\\
\phi_2^{fd(adj)}
\end{array}\right)
(y+L)=\e^{i\beta\sigma_3}
\left(
\begin{array}{c}
\phi_1^{fd(adj)}\\
\phi_2^{fd(adj)}
\end{array}\right)
(y),
\end{equation}
where $y$ stands for the coordinate of the $S^1$.
\par
The effective potential is recast as 
\begin{eqnarray}
{\bar V}_{eff}
&=& V_{eff}/ ({{3\over {4\pi^2L^5}}})\nonumber \\
&=&
4\sum_{n=1}^{\infty}{1\over n^5}
[-1+D(z_g, n)\cos(n\beta)]
\sum_{1\leq i < j \leq N}2\cos[n(\theta_i -\theta_j)]\nonumber\\
&+&
4N^{adj}\sum_{n=1}^{\infty}{1\over n^5}
[1-D(z_{adj}, n)\cos(n\beta)]
\sum_{1\leq i < j \leq N}2\cos[n(\theta_i -\theta_j)]\nonumber\\
&+&
(2\times 4N^{fd})\sum_{n=1}^{\infty}{1\over n^5}
[1-D(z_{fd}, n)\cos(n\beta)]\sum_{i=1}^{N}\cos(n\theta_i).
\label{potential}
\end{eqnarray}
As seen from Eq. (\ref{potential}), supersymmetry is broken by
the bare mass explicitly even if $\beta=0$ and is restored to yield
the vanishing effective potential if we take the massless 
limit $z_i\rightarrow 0$. Since we are interested in the effect of
bare mass on the gauge symmetry breaking patterns, we take
$\beta=0$ hereafter.
\par
Let us note that the vacuum expectation values of $\Sigma$ 
and $\phi_i$ are also 
order parameters for the gauge symmetry breaking, and one should 
take the quantum correction 
to $\vev{\Sigma}$ and $\vev{\phi_i}$ into account. Essentially, there are 
two kinds of order 
parameters for the gauge symmetry breaking in the theory, one is the
component gauge field for the $S^1$ direction and the other one is 
the vacuum expectation values for the fields $\Sigma$ and $\phi_i$. 
The vacuum expectation values of $A_y$ has a periodicity of $2\pi$,
reflecting the five dimensional gauge invariance \cite{hosotania}. 
On the other hand, there is no periodicity 
for $\vev{\Sigma} (\vev{\phi_i})$. Here we
focus on the dynamics of the Wilson line phase alone,
so that we assume $\vev\Sigma=\vev{\phi_i}=0$. We will report the study
of the effective potential by 
taking both $\vev{A_y}$ and $\vev{\Sigma}$ into account \cite{habac}. 
\subsection{Gauge symmetry breaking patterns}
The parameters of the model are
\begin{equation}
(N_{adj},~~N_{fd},~~z_g,~~z_{adj},~~z_{fd}).
\end{equation}
Once we fix these parameters, we can dynamically determine the values of 
$\theta_i$'s by minimizing the effective potential (\ref{potential}). 
We take the gauge group $SU(3)$ and study the gauge symmetry breaking 
patterns of the model. 
\par
There are many free parameters in the model and it makes
the analyses of the
vacuum structure complicated. We fix the 
parameters $N_{adj}, N_{fd},z_{g}$ and $z_{fd}$, and 
$z_{adj}$ is taken to be a free parameter. We find the
minimum of the effective potential for various values of $z_{adj}$. The 
residual gauge symmetry is given by the generators 
of $SU(3)$ commuting with the Wilson line,
\begin{equation}
W={\cal P}\mbox{exp}(ig\oint_{S^1}\vev{A_y})
\label{wilson}
\end{equation}
evaluated at the vacuum configuration.
\par
Before we go to the numerical analyses, let us study the asymptotic 
behavior of the effective potential with 
respect to $z_{adj}$. If we take the limit of $z_{adj}\rightarrow 0$, the 
contribution from the hypermultiplet belonging to the adjoint
representation under the gauge group vanishes due to the cancellation
between the boson and fermion in the multiplet. Then, the
potential is governed by the contributions from the vectormultiplet and 
hypermultiplet in the fundamental representation. In this case, the vacuum
configuration is given by
\begin{equation}
(\theta_1,~~\theta_2)=({{2\pi} \over 3},~~{{2\pi} \over 3}),
\end{equation}
for which the Wilson line (\ref{wilson}) is 
proportional to the $3\times 3$ unit
matrix, so that the residual gauge symmetry is $SU(3)$. The 
gauge symmetry is not broken in the limit. 
\par
On the other hand, if we take the 
limit of $z_{adj}\rightarrow \mbox{large}$, the 
contribution from the adjoint squark to the effective
potential is decoupled 
due to the Boltzmann like suppression factor 
in Eq.(\ref{bessel}). Then, the superpartner of the
adjoint squark, the adjoint fermion contributes to the
potential and enforces the system toward the phase in which the
gauge symmetry is maximally 
broken, {\it i.e.} $U(1)\times U(1)$ for the present case. 
These observations suggest that the vacuum configuration changes 
according to the change of the values of $z_{adj}$ and that there exists 
a certain critical values for $z_{adj}$. The realized gauge symmetry
is different above or below the critical values.
\par
Let us note that the effective 
potential (\ref{potential}) is symmetric under the permutation of the
phases $\theta_i$'s, which comes from the Weyl group, and the
change of the signs of the phases. In addition, the phases are
modules of $2\pi$. Then, the fundamental 
region on $\theta_2$-$\theta_1$ plane is given by 
\begin{equation} 
0<\theta_2\leq \theta_1\qquad \mbox{and}\qquad \theta_2\leq 2\pi -2\theta_1.
\end{equation}
Once we obtain the vacuum configuration $(\theta_1, \theta_2,
\theta_3)$, where $\theta_3=-\theta_1-\theta_2$ in the fundamental 
region given above, the vacuum configurations for other regions are
obtained by the permutation of $\theta_i$'s and by taking a module of
$2\pi$ of the phase into account,
\begin{equation}
(\theta_3, \theta_1, \theta_2),~(\theta_3, \theta_2, \theta_1),~
(\theta_2, \theta_3, \theta_1),~(\theta_1, \theta_3, \theta_2),~
(\theta_2, \theta_1, \theta_3).
\end{equation}
\subsubsection{Example I} 
Let us first choose the parameters as
\begin{equation}
(N_{adj},~~N_{fd},~~z_g,~~z_{fd})=(2,~~1,~~1.0,~~0.5).
\label{seta}
\end{equation}
The vacuum configuration of the model is obtained by 
minimizing the effective potential numerically. In Fig.$1$, we depict
the location of the vacuum configuration with respect to the
change of $z_{adj}$ on 
the $x$ - $y$ plane, where $x\equiv \theta_1/2\pi, y\equiv
\theta_2/2\pi$ (mod $1$).
\par
We observe that
\par\noindent
(i)~for $0< z_{adj}\leq 0.630175$, the 
configuration $(x, y)=(1/3, 1/3)$ is the vacuum 
configuration, for which the $SU(3)$ gauge symmetry is not broken, 
\\[0.3cm]
(ii)~as $z_{adj}$ becomes larger than $0.631075$, the
configuration $(x, y)=(1/3, 1/3)$ starts to
move toward the $x$-axis. The vacuum configurations for a fixed values of
$z_{adj}$ respects $U(1)\times U(1)$ gauge symmetry,
\\[0.3cm]
(iii)~for $z_{adj}=0.75$, the vacuum configuration 
almost reach to the $x$-axis, and for larger values 
of $z_{adj}=0.75$, the vacuum configuration moves on the
$x$-axis and finally approaches to $(1/3, 0)$. As an example, if we 
take $z_{adj}=100$, the vacuum configurations are numerically 
given by $(x, y)=(0.334805, 0)$, which are close to the
configurations $(x, y)=(1/3, 0)$.
The residual gauge symmetry is given by $U(1)\times U(1)$.
\par
We summarize that 
\begin{equation}
\mbox{gauge symmetry breaking patterns}
=\left\{\begin{array}{ll}
0<z_{adj}\leq z_* & \cdots SU(3)\rightarrow SU(3),\\
z_{adj}>z_*       & \cdots SU(3)\rightarrow U(1)\times U(1),\\
\end{array}\right.
\label{patterna}
\end{equation}
where $z_*=0.630175$. We do not have the partial gauge symmetry
breaking $SU(3)\rightarrow SU(2)\times U(1)$ for the 
parameter set (\ref{seta}). We also note that the 
change of the vacuum configuration is continuous
and smooth, so that the phase transition, $SU(3)\rightarrow
U(1)\times U(1)$ is
the second order.
\begin{figure}
\centering
\leavevmode
\includegraphics[width=10.cm]{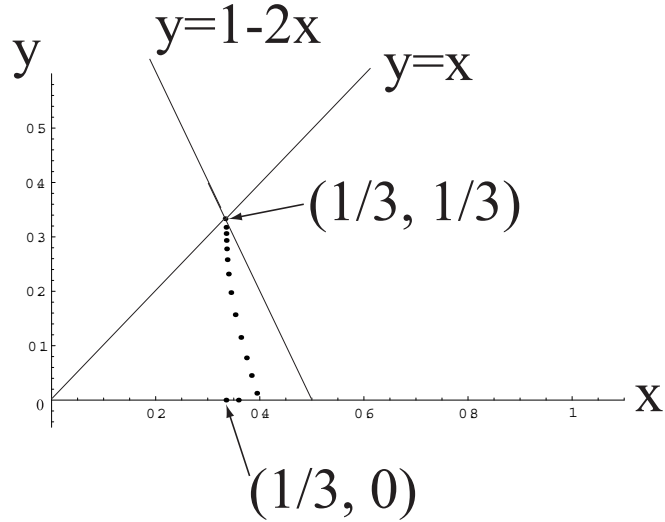}
\caption{The location of the vacuum configuration with respect to
the change of $z_{adj}$ for the parameter set (\ref{seta}). Both
axises are normalized as $x=\theta_1/2\pi, y=\theta_2/2\pi$.} 
\end{figure}
\par
\subsubsection{Example II}
We have seen that the vacuum configuration is affected by
the bare
mass and depending on the values of $z_{adj}$, the residual gauge
symmetry is different. In our previous example, it is given by
either $SU(3)$ or $U(1)\times U(1)$, and we do not have the partial gauge
symmetry breaking. Here we will show that such 
the partial gauge symmetry breaking, which is 
important in connection with GUT, is possible for an appropriate choice 
of the parameter. 
\par
Let us choose the parameter as 
\begin{equation}
(N_{adj},~~N_{fd},~~z_g,~~z_{fd})=(2,~~1,~~1.0,~~10.0).
\label{setb}
\end{equation}
We note that the number of flavors is the same as before, but
the magnitude of the bare mass, namely, $z_{fd}$ is changed.
We study the vacuum structure of the model according to the
change of the values of $z_{adj}$.  Again the vacuum configuration 
is obtained by
minimizing the effective potential numerically. In Fig.$2$, we 
depict the location of the vacuum configuration with respect to $z_{adj}$ on 
the $x$ - $y$ plane.
\par
We observe that
\par\noindent
(i)~for $0\leq z_{adj}\leq 0.501577$, the vacuum configuration is given by
$(x, y)=(1/3, 1/3)$, for which the residual gauge symmetry is $SU(3)$, 
\\[0.3cm]
(ii)~if $z_{adj}$ becomes larger than $0.501577$, the 
configuration starts to move on the 
line $y=1-2x$ toward $(x, y)=(1/2, 0)$. The residual gauge symmetry for
the configurations on the line is $SU(2)\times U(1)$. At $z_{adj}
\simeq 0.881952$, the vacuum configuration is located 
at $(x, y)=(1/2, 0)$ and stays 
there for $0.881952 \leq z_{fd}\leq 1.11002$,    
\\[0.3cm]
(iii)~for larger values of $z_{adj}\simeq 1.11002$, the 
configuration $(1/2, 0)$ moves on the $x$-axis and finally arrives 
at $(1/3, 0)$ for $z_{adj}\rightarrow \infty$. The residual gauge 
symmetry for $z_{adj}> 1.11002$ is $U(1)\times U(1)$.
\par
\begin{figure}[tbh]
\centering
\leavevmode
\includegraphics[width=10.cm]{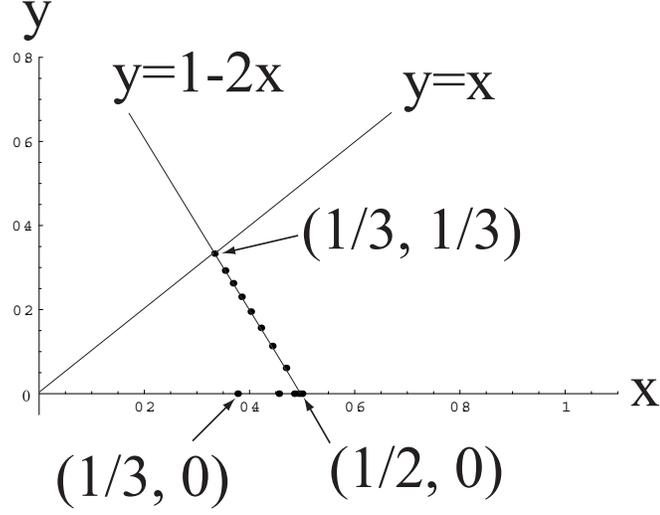}
\caption{The location of the vacuum configuration with respect to
the change of $z_{adj}$ for the parameter set (\ref{setb}). Both
axises are normalized as $x=\theta_1/2\pi, y=\theta_2/2\pi$.}
\end{figure}
\par
We summarize that
\begin{equation}
\mbox{gauge symmetry breaking patterns}
=\left\{\begin{array}{ll}
0<z_{adj}\leq z_{*1} & \cdots SU(3)\rightarrow SU(3),\\
z_{*1}<z_{adj}\leq z_{*2}    & \cdots SU(3)\rightarrow SU(2)\times U(1),\\
z_{adj}> z_{*2} & \cdots SU(3)\rightarrow U(1)\times U(1),
\end{array}\right.
\label{patternb}
\end{equation}
where $z_{*1(2)}=0.501577, 1.11002$. We also note that the 
change of the vacuum configuration is continuous
and smooth, so that the order of the phase 
transition is the second order.
\par
The vacuum configuration on the line $\theta_2=2\pi -2\theta_1$ respects
the $SU(2)\times U(1)$ gauge symmetry. This is because 
for $\theta_2=2\pi -2\theta_1$ 
\begin{eqnarray}
W&=&\mbox{diag}(\e^{i\theta_1}, \e^{i\theta_2}, \e^{-i(\theta_1+\theta_2)})
=\mbox{diag}(\e^{i\theta_1},\e^{2i\pi
-2i\theta_1},\e^{i\theta_1-2i\pi})
\nonumber\\
&=&\mbox{diag}(\e^{i\theta_1},\e^{-2i\theta_1}, \e^{i\theta_1})
= \mbox{diag}(\e^{i\theta_1},\e^{i\theta_1}, \e^{-2i\theta_1}).
\end{eqnarray}
In the last equality we have made the permutation of  
the second and third diagonal element. The generators of $SU(3)$ commuting 
with the above $W$ forms the $SU(2)\times U(1)$. We also observe the
configuration on the line $\theta_2=\theta_1$ is equivalent to the one
on the line $\theta_2=2\pi -2\theta_1$. 
\par
Let us comment on the scale of the mass term for $A_y$. 
The zero mode of $A_y$ acquires the mass term, which is evaluated 
from the second derivative of the effective potential 
at the vacuum configuration. We obtain the effective 
potential (\ref{potential}) thanks to the supersymmetry breaking due 
to the bare mass term. We notice that the five-dimensional gauge 
invariance and supersymmetry forbid the mass term for $A_y$. 
This means that the mass scale for $A_y$ 
must be the one, at which supersymmetry is broken and there is no 
local gauge invariance in five dimensions. In fact the 
mass term for $A_y$ is estimated, aside from numerical
constants, as  
\begin{equation}
m_{A_y}^2 \sim 
N_{deg}N_f\times {\bar g}^2 \times
\left\{\begin{array}{ll}
L^{-2} &\mbox{for} \quad ML > 1,\\
L^{-2} &\mbox{for} \quad ML\sim 1,\\
M^2 &\mbox{for} \quad ML\ll1,
\end{array}\right.
\end{equation} 
where ${\bar g}\equiv g/\sqrt{L}$ is the gauge coupling in four 
dimensions and $N_{deg}, N_f$ are the
on-shell degrees of freedom, the number of flavors, respectively. 
$M$ stands for the bare mass $M\equiv m_{g, adj, fd}$. 
We also observe that, as expected, the original
supersymmetry protects the mass term for $A_y$ against 
the large quantum correction of $O(L^{-1})$ for the case of $M \ll
L^{-1}$. Let us note that the mass is generated through 
the quantum correction at one-loop level.
\section{Conclusions and Discussions}
We have studied the gauge symmetry breaking patterns through the
Hosotani mechanism in the supersymmetric gauge models in 
five dimensions.
We have introduced the bare mass term for the matter field
to break supersymmetry instead of the Scherk-Schwarz mechanism of
supersymmetry breaking. And we have studied the effect of the
bare mass on the gauge symmetry breaking patterns.
\par
In order to study the vacuum structure of 
the model, we have fixed some of the parameters because the 
model contains many parameters. In the
paper, we have studied the vacuum structure by changing the values of 
$z_{adj}$ for the fixed parameter sets given 
by (\ref{seta}) and (\ref{setb}). 
As demonstrated in the text, the magnitude of the bare mass 
actually affects the vacuum structure of the model 
and changes the gauge symmetry breaking patterns.
\par
We have given two examples, in which the residual gauge symmetry
is different. For a parameter set given 
by (\ref{seta}), the gauge 
symmetry breaking patterns is (\ref{patterna}). On the 
other hand, for the parameter
set (\ref{setb}), it is given by (\ref{patternb}) and we have the
partial gauge symmetry breaking $SU(3)\rightarrow SU(2)\times U(1)$. 
The partial gauge symmetry breaking
has been found in the models with only massless 
particles \cite{hatanaka}. It should be noted that 
the partial gauge symmetry breaking is also realized 
by the appropriate choice of the magnitude of the bare mass.
It is interesting to investigate the partial gauge symmetry
breaking $SU(5)\rightarrow SU(3)\times SU(2)\times U(1)$, which
is important in connection with GUT. We also mention that the order 
parameter changes 
smoothly according to the change
of $z_{adj}$ in our example, so that the phase 
transition is the second order.
\par
As far as our numerical analyses are concerned, it seems that the 
hierarchical structure of the bare mass $z_{g} < z_{fd}$ may be 
essential for the partial gauge symmetry breaking
for the present number of flavors. Needless to say, one needs much 
more study in order to confirm the statement.
\par
Let us note that there is another way to realize 
the partial gauge symmetry breaking. In this paper we have considered
only the field satisfying the periodic boundary condition. One can
introduce the field that satisfy the antiperiodic boundary 
condition, keeping the singlevaluedness of the Lagrangian density. 
The mass spectrum is, then, modified as 
\begin{equation}
{{2\pi}\over L}(n+{\theta_i\over {2\pi}})\rightarrow 
{{2\pi}\over L}(n+{{\theta_i-\pi}\over {2\pi}}),
\end{equation}
so that, the contribution from the field to the effective 
potential is given by 
\begin{equation}
\bar V_{eff}^{anti}=
4N_{fd}^{anti}\sum_{n=1}^{\infty}{1\over n^5}
[2-2D(z_{fd}^{anti}, n)]\sum_{i=1}^{N}
\cos[n(\theta_i-\pi)].
\end{equation}
If we take account of the field satisfying the antiperiodic boundary
condition and choose the parameters as 
\begin{equation}
(N_{adj},~~N_{fd},~~N^{anti}_{fd}~~z_g,~~z_{fd},~~z_{fd}^{anti})
=(0,~~1,~~1~~,0.0,~~0.5,~~0.5)
\label{setc}
\end{equation}
for example, we can see $(\theta_1,\theta_2)=(\pi/3, \pi/3)$
is a vacuum configuration \footnote{Although $(\theta_1,\theta_2)
=(2\pi/3,2\pi/3)$ is also a vacuum for this parameter set
(\ref{setc}), a little bit smaller $z_{fd}$, {\it e.g.}
$z_{fd}=0.4$ makes a vacuum that realize 
the partial gauge symmetry breaking the
global minimum.}. This shows that 
in addition to the magnitude of the 
bare mass, the periodicity of the matter field also changes 
the gauge symmetry breaking patterns. This is an 
interesting problem to study further.
\par
It is also important to study the effect of the bare mass on the gauge
symmetry breaking patterns in case of orbifold, for instance, $S^1/Z_2$. In
particular, it is interesting to study how the bare mass affects the
mass of the Higgs scalar embedded in $A_y$ as 
studied in \cite{haba} \cite{habab}. 
\par
\begin{center}
{\bf Acknowledgements}
\end{center}  
N.H is supported in part by the Grant-in-Aid for Science
Research, Ministry of Education, Science and Culture, Japan,
No.~16540258, 16028214, 14740164. K.T would thank the colleagues 
in Osaka University and the professor Y. Hosotani for 
valuable discussions. K.T is supported by the $21$st Century 
COE Program at Osaka University. T.Y thanks the Japan Society
for the Promotion of Science for financial support. 
\newpage

\end{document}